\documentclass[%
 reprint,
superscriptaddress,
 amsmath,amssymb,
 prl,aps,
]{revtex4-1}

\usepackage{graphicx}
\usepackage{dcolumn}
\usepackage{bm}
\usepackage{hyperref}
\usepackage{xargs}                      
\usepackage{xcolor,ulem}  
\usepackage[colorinlistoftodos,prependcaption,textsize=tiny]{todonotes}
\newcommandx{\unsure}[2][1=]{\todo[linecolor=red,backgroundcolor=red!25,bordercolor=red,#1]{#2}}
\newcommandx{\change}[2][1=]{\todo[linecolor=blue,backgroundcolor=blue!25,bordercolor=blue,#1]{#2}}
\newcommandx{\info}[2][1=]{\todo[linecolor=OliveGreen,backgroundcolor=OliveGreen!25,bordercolor=OliveGreen,#1]{#2}}
\newcommandx{\improve}[2][1=]{\todo[linecolor=green,backgroundcolor=green!25,bordercolor=green,#1]{#2}}
\newcommandx{\thiswillnotshow}[2][1=]{\todo[disable,#1]{#2}}

\newcommand{\angstrom}{\mbox{\normalfont\AA} }

\begin{document}

\preprint{APS/123-QED}

\title{First-principles theory of giant Rashba-like spin-splitting in bulk ferroelectrics}
\author{Louis Ponet}
\email{louis.ponet@iit.it}
\affiliation{%
 Quantum Materials Theory, Istituto Italiano di Tecnologia, Genoa 16163, Italy
}
\affiliation{Department of Nanosciences, Scuola Normale Superiore di Pisa 56100, Italy}
\author{S. Artyukhin}%
 \email{sergey.artyukhin@iit.it}
\affiliation{%
 Quantum Materials Theory, Istituto Italiano di Tecnologia, Genoa 16163, Italy
}
\date{\today}

\begin{abstract}
Recently large Rashba-like spin splitting has been observed in certain bulk ferroelectrics. In contrast with the relativistic Rashba effect, the chiral spin texture and large spin-splitting of the electronic bands depend strongly on the character of the band and atomic spin-orbit coupling.
We establish that this can be traced back to the so-called orbital Rashba effect, also in the bulk.
This leads to an additional dependence on the orbital composition of the bands, which is crucial for a complete picture of the effect. Results from first-principles calculations on ferroelectic GeTe verify the key predictions of the model.
\end{abstract}

\pacs{Valid PACS appear here}
\maketitle

Bulk ferroelectrics with large atomic spin-orbit coupling allow for electric control of spin-polarized states~\cite{DiSante2013,Ishizaka2011,Kim2014, Liebmann2016, Krempasky2015SurfaceSemiconductor}, allowing for the switching of the spin texture by an externally applied electric field. 
The underlying mechanism is, however, not well understood.

At first glance the splitting appears to be an abnormally large Rashba-effect, which seems plausible given the presence of a non-zero electric polarization along the $z$-axis. However, if one considers the relativistic Rashba Hamiltonian \cite{Rashba1959SymmetryAr,Lifshitz1982CourseTheory}
\begin{equation}
\mathcal{H}_R=\alpha_R (\bm{k} \times \bm{E})\cdot \bm{\sigma}, \quad \alpha_R=\frac{e\hbar^2}{2m^2c^2}
\end{equation}
where the electric field originates from the non-zero polarization $\bm{E}$, an enormously big value of $\alpha_R\approx$ 30.7 eV$.$~\angstrom would have to be used \cite{DiSante2013} to approximate the band structure. In vacuum this relativistic constant is of the order of $10^{-6}$ eV. \angstrom. Another issue, as has been experimentally confirmed\cite{Krempasky2015SurfaceSemiconductor}, is that the orientation of the spin polarization of the spin-split sub-bands depends on the atomic orbitals that form the band. 
This is not accounted for by the relativistic Rashba effect, where the orientation of spin-polarization is uniquely defined by the vector product of the wave vector and the inversion symmetry breaking field (in our case the electric polarization).
The last discrepancy is that the splitting depends strongly on the atomic SOC.
It is then evident that the mechanism behind the spin-splitting must be one that couples the atomic character of the bands (and their respective atomic SOC) to the electric polarization.

This work aims to clarify the effect in {\it bulk} systems, the interaction between ferroelectricity and atomic SOC that cause it, and in doing so identifies the criteria for finding other materials with large effect.

 We will first present a short overview of the important properties of GeTe,
followed by a qualitative description of the coupling between the orbital angular momentum and the electric polarization. Then we verify the model by performing ab-initio calculations on Germanium Telluride (GeTe), a typical example of the aforementioned effects.

\begin{figure}[ht!]
\includegraphics[width=\linewidth]{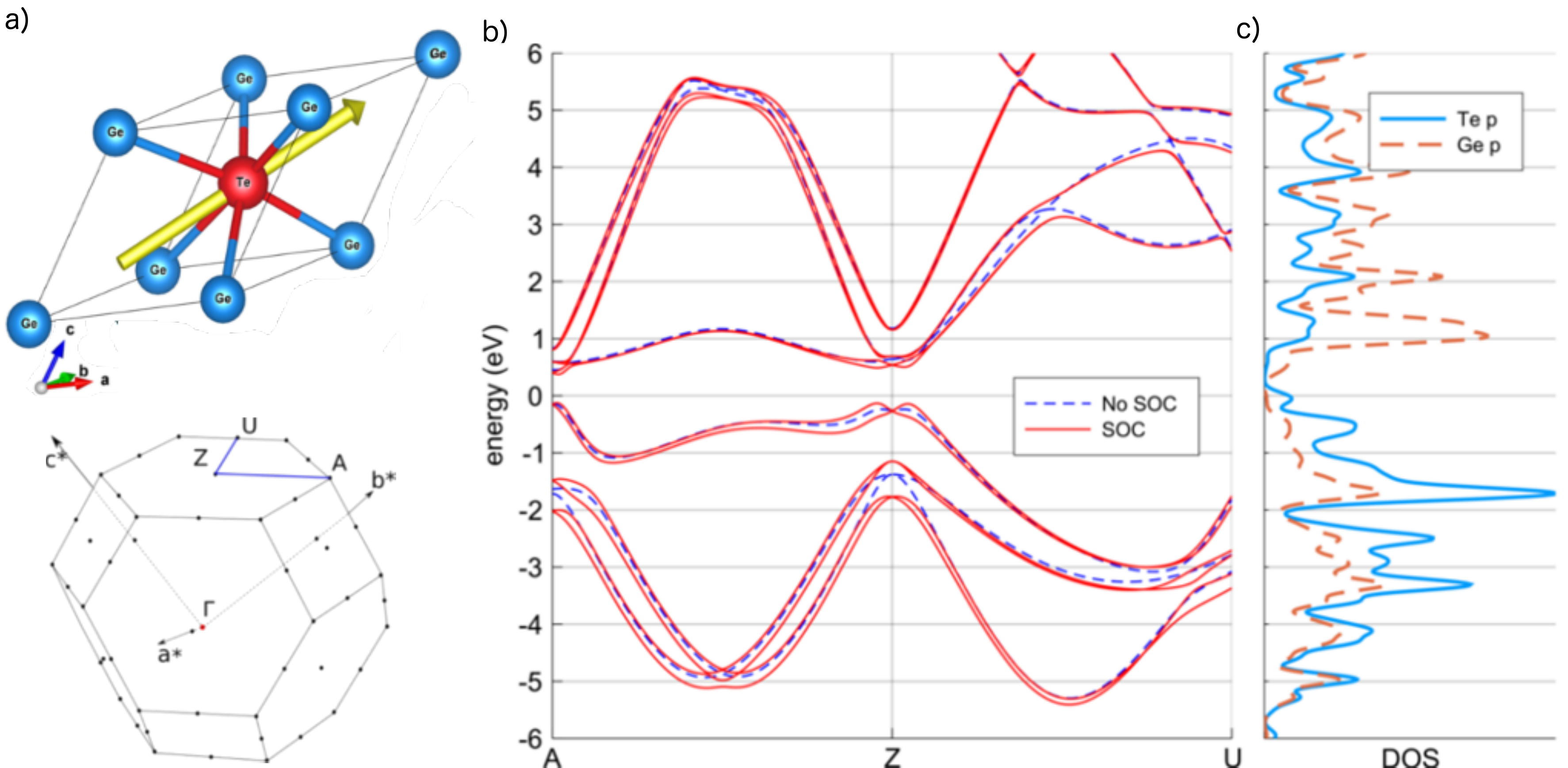}
\caption{\label{fig:eigvalsdos}a) Rhombohedral unit cell and Brillouin zone of GeTe, with the polarization direction in yellow. b) Band structure obtained from a DFT calculation with and without SOC, along the blue path in panel a). c) partial DOS for Te and Ge p-orbitals computed without SOC.}
\end{figure}

Germanium Telluride (GeTe) is a ferroelectric semiconductor with R3m space group. The ferroelectricity is owed to a small displacement along the threefold rotation axis ($z$) of the Te layers towards one of the two neighboring Ge layers \cite{Rabe1987}. The resulting electric polarization is therefore also oriented along the $z$-axis, as can be seen in Fig. \ref{fig:eigvalsdos} (a). The bandstructure, displayed in Fig. \ref{fig:eigvalsdos} (b), presents a distinct large linear spin-splitting around the Z point, resembling the well known Rashba-effect \cite{DiSante2013}. The characteristic splitting happens both along the Z-A path and Z-U path, but not along Z-$\Gamma$ path, due to time-reversal symmetry.

The density of states (DOS) is displayed in Fig.~\ref{fig:eigvalsdos} (c). The valence bands are comprised mostly of Te 5p-orbitals, whereas the conduction bands are mostly formed by Ge 4p-orbitals. This orbital character of valence and conduction bands, together with the stronger atomic SOC on Te atom, results in a more pronounced spin-splitting in the valence bands. We will focus on the first three valence bands, the top one of which mainly has $p_z$ character whereas the two lower bands, which are degenerate at the $Z$-point, are formed by $p_x$ and $p_y$ orbitals. This splitting between $p_z$ and $p_{x,y}$ is due to the distortion of the ideal Te-Ge$_6$ octahedron and the resulting crystal potential.

In their seminal papers Park et al.~addressed large Rashba-like spin splitting at surfaces, emphasizing the pivotal role of OAM \cite{Park2011,Park2012,Park2015MicroscopicMomentum,Hong2015QuantitativeSplitting}.
This is because the interference between neighboring atomic orbitals with non-zero OAM in the Bloch function can result in $k$-dependent charge asymmetry.
The resulting electric dipole couples to the inversion symmetry-breaking field at the surface, resulting in the splitting of OAM states, linear in $k$.
In the bulk inversion symmetry can be broken by e.g. electric polarization or an external electric field.

\begin{figure}[t]
~\centering
\includegraphics[width=\linewidth]{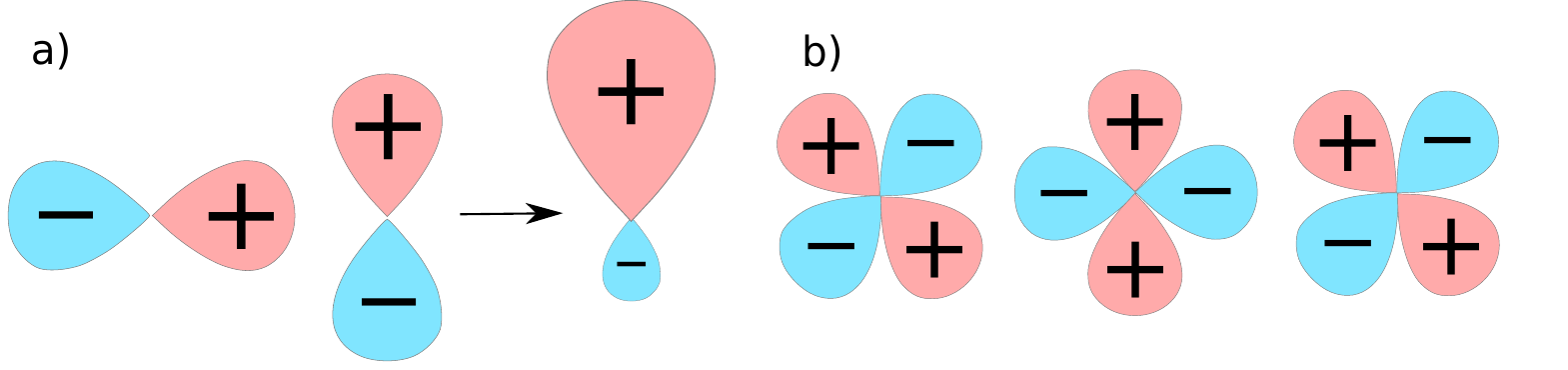}\caption{\label{fig:overlapdip} Overlap dipoles of orbitals in neighboring unit cells. a) Nonzero dipole coming from shifted $p$-orbitals. b) Dipoles of shifted $d$-orbitals compensate.}
\end{figure}

Since the operators of interest ($\hat{r}$ and $\hat{l}$) are in real space, we adopt the Wannier representation\cite{Marzari2012MaximallyApplications} to derive the microscopic Hamiltonian.
To this end we write the Bloch functions as linear combinations of the Wannier functions that describe the bands of interest: $\psi_{\bm{k}}(\bm{r})=\sum_{\alpha,\bm{R}} c_\alpha(\bm{k}) w_\alpha(\bm{r}-\bm{R})e^{i\bm{k}\bm{R}}$, where the $w_\alpha$ are chosen to be real and $\bm{R}$ denote the unit cell in which the orbital is centered.

In the case of GeTe the electric polarization is along the z-direction, making it sufficient to focus on the corresponding dipole moment
\begin{align}
d_z(\bm{k}) &=e\sum_{\alpha,\beta,\bm{R}}A_{\alpha,\beta}(\bm{k})e^{i\bm{k}.\bm{R}} \, \mathcal{Z}_{\alpha,\beta}^{\bm{R}} \\
A_{\alpha,\beta}(\bm{k}) &= c_{\alpha}^*(\bm{k}) c_\beta(\bm{k}) \\
\mathcal{Z}_{\alpha,\beta}^{\bm{R}} &= \int d\bm{r}   w_{\alpha}(\bm{r})w_\beta(\bm{r}-\bm{R}) z \label{eq:Znn}
\end{align}
One can then expand the $k$-dependent variables around $k_Z$, keeping terms up to first order
\begin{align}\label{eq:dip}
d_z(\bm{k}) =e\sum_{\alpha,\beta,\bm{R}} &\left(A_{\alpha,\beta}(Z) + \bm{k}^r \left.\frac{\partial A_{\alpha,\beta}(\bm{k}) }{\partial \bm{k}}\right\rvert_{\bm{k}=Z}\right) \times\\ 
&(1 + i\bm{k}^r \cdot \bm{R}) \mathcal{Z}_{\alpha,\beta}^{\bm{R}},
\end{align}
where $\bm{k}^r = (k_Z - \bm{k})$.
The term that combines the first order variation of $A_{\alpha,\beta}(\bm{k})$ with the zeroth order of the exponent leads to a nonzero contribution only if $w_\alpha, w_\beta = s, p_z$ in the same unit cell ($\bm{R}=0$).
This is the charge asymmetry that comes from $s$-$p_z$ hybridization and was previously considered for surfaces in Ref.\cite{Petersen2000SimpleStates} and for bulk perovskites in \cite{Kim2014}.
The new terms that we consider here combine the zeroth and first order terms of $A_{\alpha,\beta}(\bm{k})$ with the first order $i\bm{k}^r \cdot \bm{R}$ term.
Assigning $1,2,3$ to $p_x, p_y, p_z$ in the usual fashion, one can find the following relation between the Bloch function coefficients and the OAM: $l_k(\bm{k}) = -i \epsilon_{ijk} A_{ij}$.

\begin{figure}[t]
~\centering
\includegraphics[width=\linewidth]{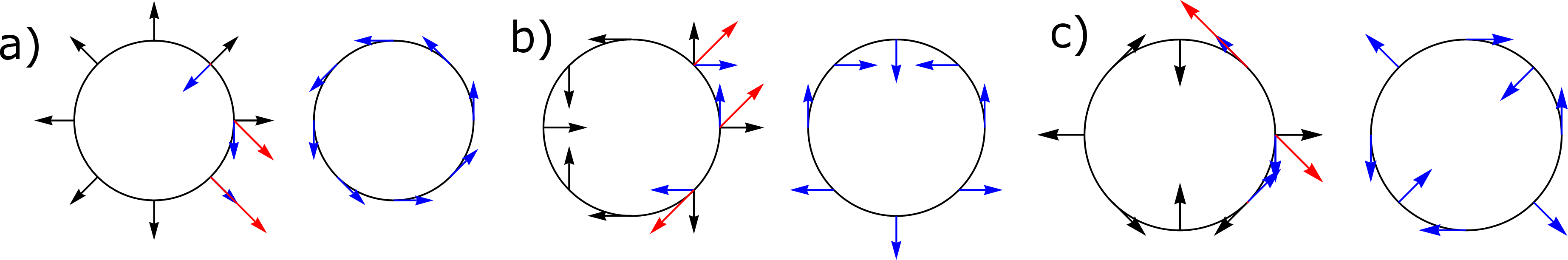}\caption{\label{fig:overlaparrows} Dipoles due to interference for $p$, $d$ and $f$ orbitals, with maximal angular momentum projection, $\psi\sim e^{i l\phi}$, are shown in panels a) - c) respectively. Black and blue arrows denote the complex phases of wavefunctions in neighboring unit cells, where the red arrow shows the resulting complex amplitude of the total Bloch function due to interference of the orbitals. The overlapping orbitals where spaced apart for visual clarity.}
\end{figure}

\begin{figure*}[t]
\includegraphics[width=0.49\textwidth]{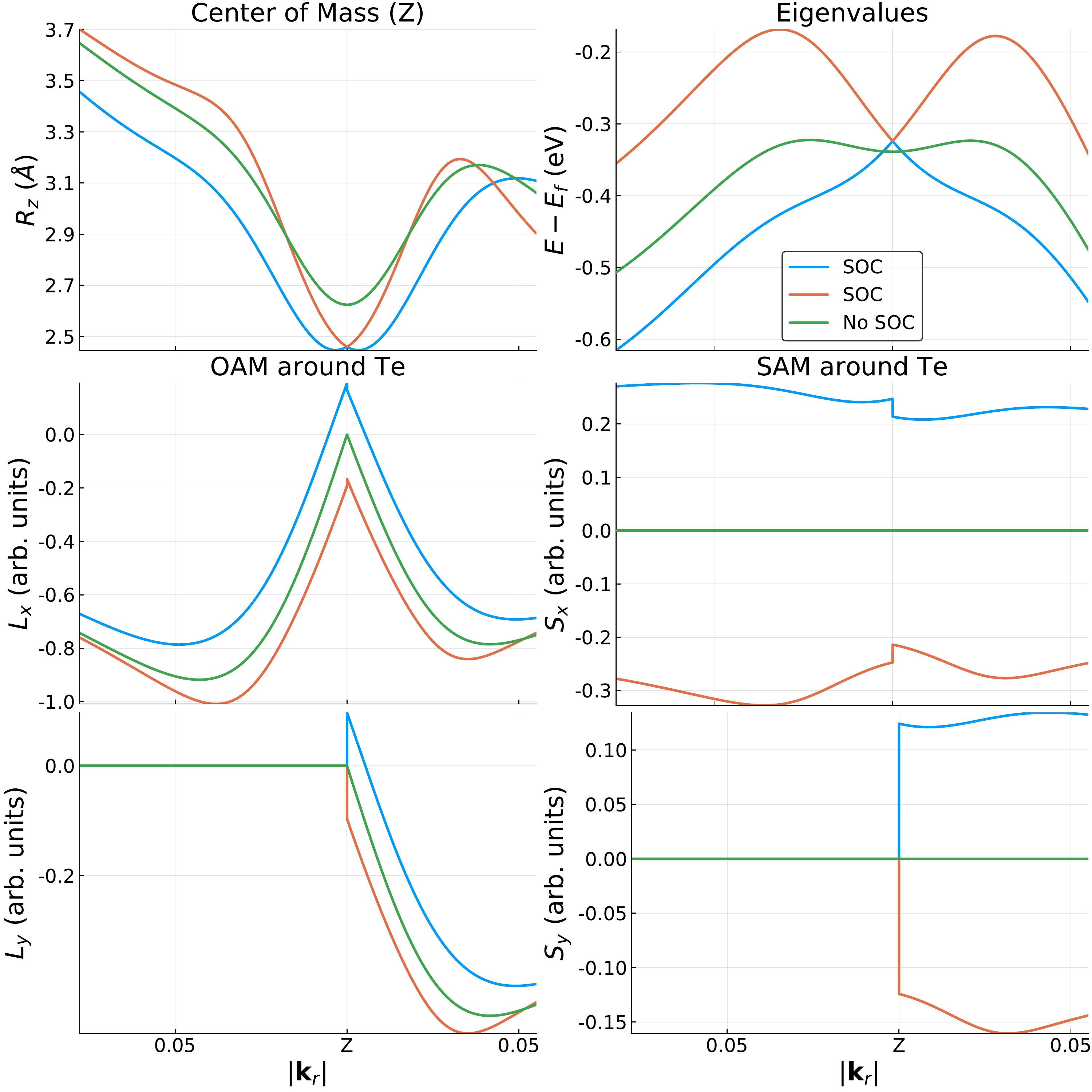}
\includegraphics[width=0.49\textwidth]{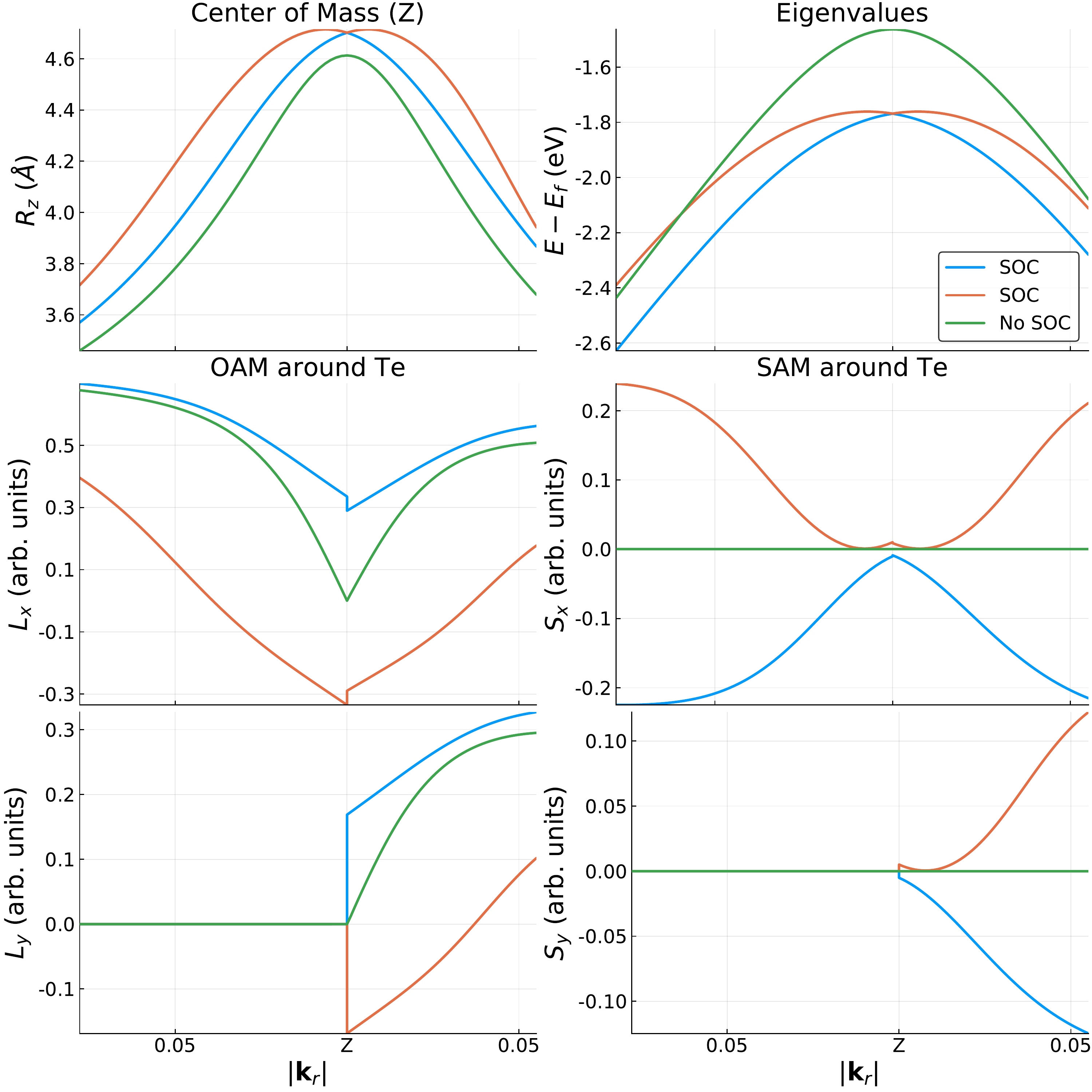}
\caption{Comparison between properties and energy dispersion in the first (left panel) and third (right panel) valence band. $\bm{k}_r = \bm{k} - \bm{k}_Z$. }\label{fig:oamvseigvalv}
\end{figure*}
Upon closer inspection of the overlap dipole, $\mathcal{Z}_{\alpha,\beta}^{\bm{R}}$, one can conclude that in order to get a dipole along $z$, at least one of the orbitals needs to be the $p_z$ orbital and the shift-vector $\bm{R}$ has to have a component along the second orbital ($p_x$ or $p_y$). This is illustrated in Fig.~\ref{fig:overlapdip}(a). Another requirement for this term to be nonzero is that the orbitals have odd parity, such as $p$ and $f$. This is highlighted in Fig.~\ref{fig:overlaparrows}. If they have even parity (as is the case for the $d$-orbitals displayed in Fig.~\ref{fig:overlapdip}(b) and ~\ref{fig:overlaparrows}(b)), the overlap dipoles from neighboring unit cells will cancel. 
Filling this into Eq.~(\ref{eq:dip}) and summing over the nearest neighbor unit cells ($\bm{R} = \pm 1$), we arrive at
\begin{align}\label{eq:dipfinal}
    d_z(\bm{k}) &= 4e \left(l_x(Z) + \bm{k}^r \left.\frac{\partial l_x(\bm{k}) }{\partial \bm{k}}\right\rvert_{\bm{k}=Z}\right) k^r_yR_y \mathcal{Z}_{y,z}^{R_y}\nonumber \\
    &  - 4e\left(l_y(Z) + \bm{k}^r \left.\frac{\partial l_y(\bm{k}) }{\partial \bm{k}}\right\rvert_{\bm{k}=Z}\right) k^r_xR_x \mathcal{Z}_{x,z}^{R_x}.
\end{align}
This term will couple to an external electric field, or in the case of GeTe, the electric polarization.
This leads to a term in the Hamiltonian of the form:
\begin{equation}
    \mathcal{H}_{OR} \propto \bm{l} \cdot ( \bm{k} \times \bm{E})
\end{equation}
The terms with $l_{x,y}(Z)$ lead to a linear $k$-dependence if the atomic SOC unquenches the OAM.
The second term leads to energy quadratic in $k$, but crucially it results in a linear variation of the OAM even without including SOC. This linear variation of the OAM will then couple to $\bm{\sigma}$ (also referred to as spin angular momentum (SAM)) through the atomic SOC, leading to the linear spin-splitting.

The effective Hamiltonian, including SOC, is given by
\begin{equation}\label{eq:hami}
\mathcal{H} = \mathcal{H}_0 + \lambda_{so} \bm{l}\cdot \bm{\sigma} + c_1 \bm{l} \cdot ( \bm{k} \times \bm{E}) + c_2 \bm{l}^2+ c_2' l_z^2,
\end{equation}
where $c_2 \bm{l}^2+ c_2' l_z^2$ is the contribution due to the crystal field splitting.
To explicitly show how the linear variation of the OAM arises from the terms in Eq.~(\ref{eq:dipfinal}), we adopt the mean-field treatment, substituting the operators for their average values. Similar results can be obtained using the Kubo formula. This leads, after minimizing the energy with respect to the OAM, to the following expressions
\begin{align}\label{eq:OAM}
l_x &= -\frac{\lambda_{so}\sigma_x - c_1 E_z k_y}{2c_2} \\
l_y &= -\frac{\lambda_{so}\sigma_y + c_1 E_z k_x}{2c_2},
\end{align}
only including terms that depend on  $E_z$.

We performed ab-initio calculations to study how the discussed effects manifest themselves in GeTe. To arrive at the desired basis of Wannier functions and tight-binding Hamiltonian, we first performed a collinear DFT calculation using the Quantum-Espresso package \cite{Giannozzi2009}, followed by WANNIER90 \cite{Mostofi2014AnFunctions}. For the DFT calculation we used a 10x10x10 Monkhorst-Pack k-grid, as well as an energy convergence threshold of $10^{-8}$~Ry. In the Wannierization step the gauge freedom of Wannier functions was exploited to arrive at a set of basis functions which are localized \cite{Marzari2012MaximallyApplications}, and resemble closely the atomic orbitals (spherical harmonics). Afterwards we added atomic SOC in the form $\lambda_{so} \bm{l} \cdot \bm{\sigma}$ using the muffin-tin approximation, with $\lambda_{so}$ for both atoms used as fitting parameters. Using this basis and Hamiltonian, we can the calculate the observables of interest, namely the OAM around Te and the dipole of the Bloch functions. 

Focusing on the topmost valence band, displayed in the left panel of Fig.~\ref{fig:oamvseigvalv}, there are several features that warrant a discussion. The first is that we clearly see the linear variation of the OAM even without SOC, with forms given by Eq.~(\ref{eq:OAM}). Secondly, since for the A-Z path of the BZ only $k_y$ varies, we can see that the OAM along the y-axis ($l_y$) vanishes along the entire path. It is only when $k_x$ varies, as the wavevector progresses along the Z-U path, that we observe a non-vanishing $l_y$. Along this path, the sign of $k_x$ is negative whereas the sign of $k_y$ is positive just as along the Z-A path. This results in a consistent orientation of $l_y$ and $l_x$, given by Eq.~(\ref{eq:OAM}). The third observation is that after including atomic SOC there is a clear manifestation of the "unquenching" of the OAM. In Fig. ~\ref{fig:oamvseigvalv}, as soon as there is an infinitesimal variation of the the wavevector from the time-reversal invariant Z-point (where $\bm{l}=0$ due to the symmetry), the unquenching leads to a nonzero shift in the OAM, depending on the sub-band.
Lastly, the variations of the dipole and OAM of the bands are correlated, consistent with Eq.~(\ref{eq:dipfinal}).
These considerations also hold true for the third valence band, as seen from the right panel of Fig.~\ref{fig:oamvseigvalv}.

There are also differences which cannot be explained by the relativistic Rashba effect, as was discussed above. For states with $j=\frac{3}{2}$ or $j=\frac{1}{2}$, the SAM is oriented along or opposite to the OAM, respectively. Comparing the orientations of the SAM and OAM shown in two panels of Fig.~\ref{fig:oamvseigvalv}, we can conclude that the topmost valence band has mostly $j=\frac{3}{2}$ character, whereas the third mostly $j=\frac{1}{2}$. 
As a whole the spin texture is opposite in the two bands.
This difference in spin texture between the two bands can only be explained by terms such as Eq.~(\ref{eq:dipfinal}). 
The relativistic Rashba effect does not share this dependence, and would thus lead to the same spin texture in each band.

Looking back to Eq.~(\ref{eq:hami}) three possible terms could result in a large linear spin-splitting energy dispersion can be identified. The first is a large unquenching effect, resulting in large $\bm{l}(Z)$ and $\bm{\sigma}(Z)$, causing a linear variation of the dipole energy, following Eq.~(\ref{eq:dipfinal}). The second is a constant OAM coupled to a linear variation of the SAM, or vice versa for the third. 
In the topmost valence band the linear variations of the dipole and SAM are very small. This suggests that the origin of the giant Rashba-like splitting is the large linear variation of the OAM, caused by the coupling to the electric polarization through the dipole, together with the large atomic SOC which then couples to the non-zero SAM through $\lambda_{so} \bm{l} \cdot \bm{\sigma}$. In the third valence band, however, the variation of the OAM is less than that of the SAM, and the contribution due to the charge asymmetry plays a bigger role, making it hard to assign the splitting to a single contribution. 

We have explored the microscopic origin of the giant Rashba-like spin-splitting in the band structure of bulk ferroelectrics with high atomic SOC. We derived a microscopic expression that results in the observed effective Hamiltonian, relating the large spin-splitting to the intricate interplay between OAM, atomic SOC, the crystal field and the electric polarization. It turns out that the crucial component, that was not considered previously, is the emergence of a nonzero electric dipole of the Bloch functions due their OAM. The quantitative analysis based on Wannier functions and muffin-tin approximation confirms this mechanism in GeTe. We find a very good agreement between the proposed effective Hamiltonian given in Eq.~(\ref{eq:hami}) and the energy dispersions of the first and third valence bands, where the effect manifests itself most clearly. 

This suggests large ferroelectric polarization, high atomic SOC and highly symmetric environment producing little $\bm{l}$ quenching by crystal field effects as design rules for new materials with strong Rashba-like spin-splitting. These could enable spintronic devices with the much needed electric control of the spin-polarization.

\end{document}